\documentclass[a4paper,review,12pt,authoryear]{elsarticle}

\usepackage{amsfonts,amsmath,bbm,bm,xcolor,booktabs,hyperref,amsthm}
\usepackage{geometry}
\usepackage{todonotes}
\usepackage{hyperref}
\usepackage{ulem}
\usepackage{subfig}
\usepackage{setspace}
\let\code=\texttt

\usepackage{enumitem}

\newcounter{researchquestion}
\newenvironment{researchquestion}{%
    \refstepcounter{researchquestion}%
    \par\medskip\noindent%
    \textbf{RQ\theresearchquestion.}~%
}{\medskip}

\hypersetup{
  colorlinks=true,
  linkcolor=blue,
  citecolor=blue,
  urlcolor=blue}

\begin{document}

\begin{frontmatter}

\title{Constructing hierarchical time series through clustering: \\Is there an optimal way for forecasting?}

  \author[label1]{Bohan Zhang\corref{cor1}}
  \address[label1]{School of Economics and Management, Beihang University, Beijing, China}
  \ead{zhangbohan@buaa.edu.cn}
  \cortext[cor1]{Corresponding author.}
  \author[label2]{Anastasios Panagiotelis}
  \address[label2]{The University of Sydney Business School, NSW 2006, Australia}
  \author[label3]{Han Li}
  \address[label3]{Department of Economics, The University of Melbourne, VIC 3010, Australia}

  \begin{abstract}

    Forecast reconciliation has attracted significant research interest in recent years, with most studies taking the hierarchy of time series as given. We extend existing work that uses time series clustering to construct hierarchies, with the goal of improving forecast accuracy, in three ways. First, we investigate multiple approaches to clustering, including not only different clustering algorithms, but also the way time series are represented and how distance between time series is defined. We find that cluster-based hierarchies lead to improvements in forecast accuracy relative to two-level hierarchies. Second, we devise an approach based on random permutation of hierarchies, keeping the structure of the hierarchy fixed, while time series are randomly allocated to clusters. In doing so, we find that improvements in forecast accuracy that accrue from using clustering do not arise from grouping together similar series but from the structure of the hierarchy. Third, we propose an approach based on averaging forecasts across hierarchies constructed using different clustering methods, that is shown to outperform any single clustering method. All analysis is carried out on two benchmark datasets and a simulated dataset. Our findings provide new insights into the role of hierarchy construction in forecast reconciliation and offer valuable guidance on forecasting practice. \\

  \end{abstract}


\end{frontmatter}

\newpage

\section{Introduction}

Applications where some variables are aggregates of one another, or so-called \textit{hierarchical time series (HTS)}, are found in many forecasting problems ranging from supply chain management (\citealp{syntetosSupplyChainForecasting2016}) to tourism planning (\citealp{kourentzesCrosstemporalCoherentForecasts2019}), electrical load forecasting (\citealp{jeonProbabilisticForecastReconciliation2019}), and retail demand forecasting (\citealp{makridakisM5AccuracyCompetition2022}). In recent decades, there has been an increasing interest in hierarchical forecasting, primarily driven by the success of the optimal reconciliation framework (\citealp{hyndmanOptimalCombinationForecasts2011,wickramasuriyaOptimalForecastReconciliation2019, panagiotelisProbabilisticForecastReconciliation2023}). The original motivation for forecast reconciliation was to ensure forecasts are \textit{coherent}, that is they respect the aggregation constraints implied by the hierarchical structure. Coherent forecasts facilitate aligned decisions by agents acting upon different variables within the hierarchy. For example, consider a retail setting, where a warehouse manager supplies stock to individual store managers within their region. Forecasts could be incoherent when the warehouse manager forecasts low total demand while store managers forecast high demand, leading to supply shortages. Numerous case studies in the literature demonstrate that reconciliation approaches not only yield coherent forecasts but also enhance overall forecast performance (\citealp{AthanasopoulosForecastReconciliationReview2023}).

A limitation in the overwhelming majority of forecast reconciliation studies is that the structure of the hierarchy is taken as \textit{given}. This structure usually includes \textit{bottom level series}, an overall aggregate or \textit{top level series}, with various aggregation schemes used to construct \textit{middle-level series}. Typically, middle levels are formed according to inherent attributes of the bottom-level series, such as geographical location, gender, product category, travel purpose, and others. We refer to this type of structure as the \textit{natural hierarchy}. While in some forecasting applications, decisions must be made with respect to the natural hierarchy, in other settings there might be some flexibility in determining how bottom levels are aggregated into middle levels. Moreover, a predefined natural hierarchy may not always exist in hierarchical settings, underscoring the importance of a hierarchical construction framework for forecasting purposes. It should be noted that very little attention has been paid to whether middle-level series can be constructed in a way that leads to further improvements in forecast accuracy relative to a given natural hierarchy. {To the best of our knowledge, only \cite{yang2017reconciling}, \cite{pangHierarchicalElectricityTime2018}, \cite{liForecastReconciliationApproach2019}, \cite{pangHierarchicalElectricityTime2022}, \cite{cini2023graph}, and \cite{matteraImprovingOutofSampleForecasts2023} have attempted to \textit{construct} middle-level series in a data-driven way which ultimately improved forecast accuracy. } However, all of these works use time series clustering to construct hierarchies in a manner that is somewhat ad hoc. Our work conducts a more thorough investigation of issues faced when  constructing hierarchical structures in forecast reconciliation. In particular, we address the following three research questions:

\begin{researchquestion} In terms of forecast performance, can the use of middle-level series lead to improvement compared to a two-level hierarchy (consisting of only top and bottom time series)? If so, is it possible to construct hierarchies in a data-driven way that leads to further improvements in forecast accuracy?
\end{researchquestion}

To investigate these questions, we consider two widely used empirical HTS datasets; the first is Australian tourism demand, the second, cause-of-death mortality data.  Throughout the paper, all evaluations are carried out using the series common to all hierarchies; namely the top and bottom level series. In both datasets, we find that the natural hierarchy outperforms the two-level hierarchy, and data-driven hierarchy via clustering can further improve forecast performance compared to natural hierarchy.

The rationale behind the data-driven approach lies in grouping time series with similar patterns together, thereby creating middle-level series with enhanced signals and consequently, improved forecastability. Such arguments have been put forward by  \cite{pangHierarchicalElectricityTime2018}, \cite{liForecastReconciliationApproach2019}, \cite{pangHierarchicalElectricityTime2022}, and \cite{matteraImprovingOutofSampleForecasts2023}. They all demonstrate superior forecast performance from hierarchies constructed via clustering, relative to the natural hierarchy or the two-level hierarchy. However, these studies for the most part focus on a small number of (in some cases, a single) clustering techniques. In this paper, we take a more systematic approach by clustering time series using different representations (the original time series, forecast errors, features of both), different distance metrics (Euclidean, dynamic time warping), and different clustering paradigms ($k$-medioids, hierarchical). The models used to obtain base forecasts and the reconciliation method are fixed throughout the experiments. Using both empirical datasets, as well as a simulation study, we find evidence that constructing hierarchies via clustering can lead to improved forecasting performance, although the optimal clustering method depends on the dataset as well as the base forecasting and reconciliation method.% \textbf{We use the same base forecasting model and reconciliation method through this paper}.

While the idea behind time series clustering is intuitively appealing, the increased accuracy when using clustering-based methods may be attributed to two factors. The first, which we refer to as     ``grouping'' is the idea that some correct subsets of series are chosen to form new middle-level series. This is the argument commonly made to support clustering-based hierarchy construction \citep[see \textit{e.g.}][]{liForecastReconciliationApproach2019, pangHierarchicalElectricityTime2022, matteraImprovingOutofSampleForecasts2023}.  
%The idea that often motivates time series clustering is these ``groupings'' be composed of similar series. 
The second factor, which we refer to as the ``structure'' of the hierarchy, includes the number of middle-level series, the depth of the hierarchy, and the distribution of group sizes in the middle layer(s). Evidence showing that clustering within a forecast reconciliation framework leads to improved forecast accuracy, does not disentangle contributions from these two factors. Indeed, clustering methods may only work in so far as they generate a larger number of base forecasts. {This argument would be consistent with the interpretation of reconciliation as a forecast combination of ``direct'' and ``indirect'' forecasts (\citealp{hollymanUnderstandingForecastReconciliation2021, di2024forecast}), since more middle-level series implies a greater number of indirect forecasts in the combination.} This leads to our second research question:

\begin{researchquestion}
    Should the improved accuracy of clustering-based methods be attributed to grouping together similar time series, or to the structure of the hierarchy? %including the number of middle-level series?
\end{researchquestion}

To investigate this question, we devise the following approach. We take a hierarchy found using a certain clustering method (or even the natural hierarchy), and then randomly permute the bottom level series (\textit{i.e.}, the leaf nodes of the hierarchical tree). Multiple new ``twin'' hierarchies are formed with an identical structure to the original hierarchy, but with permuted leaves. In this way, we keep the \text{hierarchical structure} fixed, but alter how series are combined. This method can be thought of as an informal ``permutation type'' test \citep{welch1990}. Our main finding is that hierarchies constructed using clustering methods do not significantly outperform their random ``twins'', leading to the conclusion that the driver of forecast improvement is the enlarged number of series in the hierarchy and/or its structure, rather than similarities between the time series.

Finally, from a practical perspective, we investigate the role of forecast combination in cluster-based hierarchical forecasting. With multiple hierarchies available and inspired by the forecast combination literature (\citealp{wangForecastCombinations50year2022}), we consider the last research question

\begin{researchquestion}
    Does an equally-weighted combination of reconciled forecasts derived from multiple hierarchies improve forecast reconciliation performance?
\end{researchquestion}

Note that forecast combination here differs from that of \cite{hollymanUnderstandingForecastReconciliation2021}, in that our approach averages not only different coherent forecasts, but also across hierarchies with completely different middle-level series. This is possible since only coherent bottom and top level forecasts are averaged and evaluated.

In summary, this paper presents four main contributions:

\begin{itemize}
  \item We introduce a novel hierarchical forecast reconciliation framework centered on hierarchy construction.  Within this framework, we introduce and compare three distinct approaches: cluster-based hierarchies, hierarchies based on random permutation, and  forecast combinations across different hierarchies.
  \item In contrast to existing literature that often focuses on a single clustering technique, our study systematically investigates the effectiveness of various time series clustering implementations. This investigation involves the incorporation of four time series representations, two distance measures, and two clustering algorithms.
  \item We conduct experiments using two empirical datasets - the Australian tourism dataset and the U.S. cause-of-death mortality dataset as well as a synthetic dataset. The results allow for a comparison of different approaches to constructing hierarchies.
  \item By constructing random hierarchies through permutation of leaf nodes, we show that the hierarchical structure is the primary contributor to improvements in forecast reconciliation performance, rather than the grouping of similar bottom level series. 

\end{itemize}

The rest of the paper is organized as follows. Section~\ref{sec:methodology} describes the trace minimization reconciliation methods employed and the clustering-based hierarchical time series augmentation techniques considered. Section~\ref{sec:cluster} first introduces the two datasets used, and then investigates RQ1, in particular the performance of cluster hierarchy compared to natural hierarchy and two-level hierarchy. Section~\ref{sec:permutation} introduces the novel permutation approach, followed by the investigation of RQ2 via evaluating the performance of natural hierarchy and the best performing cluster hierarchy compared to their respective random twins. To avoid the concern that clusters found in the empirical datasets are spurious, a simulation study is considered in Section~\ref{sec:simulation}. Section~\ref{sec:combination} covers the forecast combination approach raised in RQ3. Section~\ref{sec:conclusion} concludes this paper with discussions on the findings and outlines future research directions. 
Data and code for reproducing the results in this paper are available at \url{https://github.com/AngelPone/project_hierarchy}.

\section{Methodology}\label{sec:methodology}

\subsection{Hierarchical forecasting and reconciliation methods}

Hierarchical data can be characterised as being made up of bottom level series and their aggregates. In general, we consider a hierarchy with $n$ time series stacked into a vector $\boldsymbol{y}_t$. Let $\boldsymbol{b}_t$ denote the $m$ bottom level series, $\boldsymbol{a}_t$ denote the top level series, and $\boldsymbol{c}_t$ denote $k$ middle-level series. The top level and bottom level will be common to all hierarchies we consider, and are augmented by middle-level series. These middle-level series can be formed according to attributes of the time series or in a data-driven fashion using time series clustering. The series are linked through an $(m+k+1)\times m$ summing matrix 
\[
  \boldsymbol{y}_t = \boldsymbol{S}\boldsymbol{b}_t = \begin{bmatrix}
    \boldsymbol{A} \\\boldsymbol{C} \\ \boldsymbol{I}_m 
  \end{bmatrix}  \boldsymbol{b}_t = \begin{bmatrix}
      {a_t} \\ \boldsymbol{c}_t \\\boldsymbol{b}_t
  \end{bmatrix},
\]
where, $\boldsymbol{C}$ consists of zeros and ones that encode the aggregation, \textit{i.e.}, $c_{ij}=1$ if bottom level series $j$ is included in middle-level series $i$, and zero otherwise. The top level series aggregates all bottom levels, \textit{i.e.}, $\boldsymbol{A} = \mathbf{1}_{1\times m}$, which is a row of ones.

For the purposes of this paper, all forecasting is carried out as a two-step process. First, so called \textit{base} forecasts are produced for all series in the hierarchy and stacked into an $n$-vector $\hat{\bm{y}}$, where subscripts are suppressed for brevity. For base forecasts, we use the Exponential Smoothing (ETS) method \citep{ForecastingExponentialSmoothing}, implemented using the \code{forecast} (\citealp{forecast}) package in {R} (\citealp{R}). Alternative methods such as ARIMA models were also considered, but this choice had very little impact on the overall conclusions. 

The base forecasts generated in this first step, do not have the property that bottom level forecasts add up to forecasts of the aggregates , \textit{i.e.}, they are \textit{incoherent}. Therefore, forecast reconciliation is used as a post-forecasting step to ensure coherence of forecasts for all series in the hierarchy. In general, reconciliation takes the form of projecting the base forecasts as
\vspace{-0.1in}
\[
    \tilde{\boldsymbol{y}} = \boldsymbol{S}(\boldsymbol{S}'\boldsymbol{W}^{-1}\boldsymbol{S})^{-1}\boldsymbol{S}'\boldsymbol{W}^{-1}\hat{\boldsymbol{y}},
  \vspace{-0.15in}
\]
where $\tilde{\boldsymbol{y}}$ are the \textit{reconciled} forecasts and $\boldsymbol{W}$ is the covariance matrix of base forecast errors. {For reconciliation, we use the MinT method \citep{wickramasuriyaOptimalForecastReconciliation2019} implemented using the \texttt{FoReco} (\citealp{FoReco}) package in R, which involves plugging in a shrinkage estimator for $\boldsymbol{W}$.}Although alternatives were considered, including approaches that assume $\boldsymbol{W}$ is diagonal or an identity matrix \citep{hyndmanOptimalCombinationForecasts2011}, this choice has little impact on our main conclusions. {Despite the importance of non-negativity in practical scenarios such as demand forecasting, our study does not enforce non-negative forecasts for two reasons. First, the proportion of negative forecasts is minimal in our applications (0.25\% in the tourism dataset and 0.2\% in the mortality dataset). Second, introducing non-negativity constraints would significantly increase the computational cost of our experiments. It should be noted that the framework proposed in our paper could be applied using any reconciliation approach, including those that guarantee non-negative forecasts.}

\subsection{Time series clustering} 
\label{sec:clustering}

{Only a handful of studies have attempted to improve forecast accuracy in a reconciliation setting by constructing middle levels of the hierarchy using time series clustering. \cite{yang2017reconciling} employ spatial $k$-means clustering methods to assign distributed photovoltaics  to their nearest transmission grid bus locations for reconciliation purposes.} \cite{pangHierarchicalElectricityTime2018} detect consumption patterns of electricity smart meter data based on X-means clustering algorithm, while \cite{pangHierarchicalElectricityTime2022} propose several alternative clustering methods to group similar electricity and solar power time series. 
\cite{liForecastReconciliationApproach2019} apply agglomerative hierarchical clustering to cause-of-death time series, and
\cite{matteraImprovingOutofSampleForecasts2023} utilize Partition Around Medoids algorithms to unveil underlying structures in stock price indexes. { \cite{cini2023graph} took an end-to-end approach from which clusters can be extracted together with hierarchical forecasts.} 
However, these studies are limited in scope as they focus on a small number of  clustering techniques. 
Inspired by the comprehensive overview of time series clustering by \cite{aghabozorgiTimeseriesClusteringDecade2015a}, we consider various approaches based on three key components, namely \textit{time series representations}, \textit{distance measures}, and \textit{clustering algorithms}.

\paragraph{\textbf{Time series representations}}

The time series representation refers to the object that acts as an input for time series clustering. 
{ Our first candidate for the representation, due to its simplicity and broad applicability, is the time series itself without any processing or additional features. We refer to this as the raw time series.}
We also consider the in-sample one-step-ahead forecast error as a representation of the time series, since a key step in MinT reconciliation is to estimate the $\boldsymbol{W}$ matrix. 
It is important to note that raw time series and in-sample error representations are standardized to eliminate the impact of scale variations. 

A potential shortcoming to using the time series or forecast error as a representation is their high dimensionality, which is equal to the sample size of the training data. To address this concern, low dimension summaries of ``features'' can be considered. Features have been used in the context { of}
time series clustering by \cite{tianoFeatTSFeaturebasedTime2021}, and for forecasting by \cite{wangUncertaintyEstimationFeaturebased2022} and \cite{ liFeaturebasedIntermittentDemand2023}. 
We consider features of both the raw time series and the in-sample forecast error as representations. After filtering out the features that are constant across all series, we select $56$ features.\footnote{The list and descriptions of features are available in the online GitHub repository.} These time series features are calculated by the \code{tsfeatures} (\citealp{tsfeatures}) package in R. 
{To the best of our knowledge, we are the first to utilize in-sample forecast error and time series features as representations in the clustering-based hierarchy construction process.\footnote{We acknowledge that in-sample forecast error and time series features have been used in the computation of base forecasts or in the reconciliation process \citep[see][]{wickramasuriyaOptimalForecastReconciliation2019}.}}

\paragraph{\textbf{Distance measures}}

All clustering algorithms we consider require a distance to be defined between the objects that act as inputs to the algorithm.
We consider two widely applied distance measures: Euclidean distance and dynamic time warping (DTW). When employing Euclidean distance, dimension reduction is necessary due to the curse of dimensionality.
To address this, we perform Principal Component Analysis (PCA), extracting the first few principal components that collectively explain at least 80\% of the variance within the representations.

Unlike Euclidean distance, DTW is not as sensitive to the curse of dimensionality (\citealp{sakoeDynamicProgrammingAlgorithm1978}). Instead of performing one-to-one point comparisons, DTW accommodates time series of varying lengths through many-to-one comparisons. This flexible approach allows for the recognition of time series with similar shapes, even in the presence of signal transformations such as shifting and/or scaling.

\paragraph{\textbf{Clustering algorithms}}
In this paper, we focus on two clustering algorithms, namely $k$-medoids and agglomerative hierarchical clustering. These algorithms are implemented using the \texttt{cluster} (\citealp{cluster}) package in R. {The $k$-medoids algorithm aims to minimize the total distance between all observations within a cluster and their respective cluster median.} This is implemented using the partitioning around medoids (PAM) method (\citealp{PartitioningMedoidsProgram1990}).
Following the recommendation of \cite{PartitioningMedoidsProgram1990}, we determine the optimal number of clusters using the average silhouette width (ASW), which is a commonly used index in cluster validation \citep[see \textit{e.g.}, ][]{shutaywi2021silhouette}.

Agglomerative hierarchical clustering initialises each observation in its own cluster, and then merges the nearest two clusters in a stepwise fashion until all observations form a single cluster. We employ Ward's linkage (\citealp{murtaghWardHierarchicalAgglomerative2014a}) which defines the nearest clusters as those that minimize the increase in within-cluster variance at each step. Applying hierarchical clustering to $m$ bottom-level series results in a binary hierarchical tree with ($2m-1$) nodes, all of which are retained as middle-level series. 

\begin{figure}[!h]
    \centering
    \includegraphics[width=0.46\textwidth]{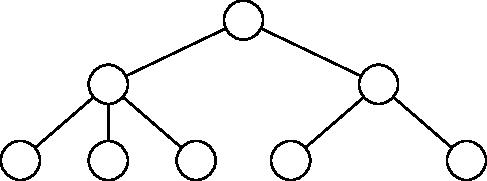}
    \hspace{1cm}
    \includegraphics[width=0.46\textwidth]{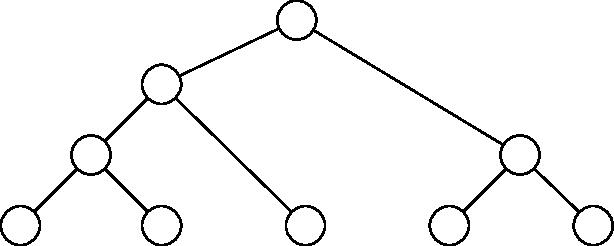}
    \caption{\label{fig:cluster_example}Example clustering results of two clustering algorithms. Left panel displays example for $k$-medoids algorithm, and right panel displays example for agglomerative hierarchical clustering algorithm.}
\end{figure}

{Figure~\ref{fig:cluster_example} illustrates two examples of hierarchies generated by $k$-medoids and agglomerative clustering algorithms, on the left and right panels respectively. The corresponding two-level hierarchy would have five bottom-level series and one top-level series. % \sout{These examples highlight the distinct behaviors of the two algorithms.}
Note that, $k$-medoids constructs a simple hierarchy with a single middle level, while hierarchical clustering generates multiple nested middle levels.} 
In general, $k$-medoids produces a hierarchy with fewer middle levels and middle-level series compared to hierarchical clustering. As the number of bottom-level series increases, these differences become increasingly pronounced, with potential implications for forecast reconciliation.
\begin{table}[!h]
\caption{\label{tab:P3_methods} Details of the 12 clustering approaches considered.}
\centering
\resizebox{\textwidth}{!}{
\begin{tabular}{rcrrr}
    \toprule
    Approach & Dimension reduction & Representation & Distance measure & Clustering algorithm  \\ \midrule
    TS-EUC-ME &  Yes & Time series  & Euclidean & $k$-medoids \\
    ER-EUC-ME & Yes& In-sample  error  & Euclidean & $k$-medoids \\
    TSF-EUC-ME & Yes& Time series features  & Euclidean & $k$-medoids \\
    ERF-EUC-ME & Yes& In-sample error features  & Euclidean & $k$-medoids \\
    TS-EUC-HC & Yes& Time series  & Euclidean & Hierarchical   \\ 
    ER-EUC-HC & Yes& In-sample error  & Euclidean & Hierarchical   \\ 
    TSF-EUC-HC & Yes& Time series features  & Euclidean & Hierarchical   \\ 
    ERF-EUC-HC & Yes& In-sample error features  & Euclidean & Hierarchical  \\
    TS-DTW-ME & No& Time series  & DTW & $k$-medoids \\
    TS-DTW-HC & No& In-sample error  & DTW & Hierarchical  \\
    ER-DTW-ME & No& Time series  & DTW & $k$-medoids \\
    ER-DTW-HC & No& In-sample error  & DTW & Hierarchical 
     \\\bottomrule
\end{tabular}}
\end{table}

In summary, we employ 12 time series clustering approaches which are derived from combinations of four time series representations, two distance measures, and two clustering algorithms. The names and details of these approaches are listed in Table~\ref{tab:P3_methods}. Note that all methods using DTW, have either the raw time series or forecast errors as representations, since DTW is not compatible with time series features. {In addition to these 12 clustering approaches, we also include a grouped hierarchy structure combining all aggregation constraints in matrix $\boldsymbol{S}$, which we refer to as ``grouped''.}

\section{Improving forecast performance via hierarchy augmentation}\label{sec:cluster}
\subsection{Data description}

We conduct our experiments on two empirical datasets throughout this paper. The first is the monthly Australian domestic tourism dataset, covering the period from January 1998 to December 2016.\footnote{Please refer to Section 4 of \cite{wickramasuriyaOptimalForecastReconciliation2019} for an in-depth explanation of this dataset.} The data is recorded as ``visitor nights'', representing the total number of nights spent by Australians away from home. In this dataset, the total visitor nights of Australia is geographically disaggregated into seven states and territories, which are further divided into $27$ zones, and then into $76$ regions. Additionally, each regional-level series is divided by four travel purposes. Overall, this dataset comprises a total of {$525$} unique time series with $304$ of those at the bottom level. In the case of tourism data, the first two or three letters of the series name indicate geographical zones or regions, and the last three denote travel purposes. For example, ``\textit{AAAHol}'' represents the visitor nights spent for holiday in the ``Sydney'' region.

The second dataset focuses on cause-of-death mortality in the U.S. We obtain monthly cause-specific death count data from the Center for Disease Control and Prevention (CDC) for the period between January 1999 and December 2019. The dataset, organized based on the 10th revision of the International Classification of Diseases (ICD) 113 Cause List\footnote{For more detailed information on the dataset, please refer to \url{https://wonder.cdc.gov/ucd-icd10-expanded.html}.}, forms a hierarchy containing $120$ time series, with $98$ of those being bottom-level series\footnote{To address the data suppression issue, we combined certain ICD codes to ensure all death counts are no less than 10. }. The top-level series represents the aggregated deaths from all causes, while the middle-level series are constructed based on major cause-of-death groups. As an example, \textit{Diseases of heart} (ICD code: I00--I09, I11, I13, I20--I51; 113 Cause List: GR113-054) is a middle-level series in the hierarchy, which contains bottom-level series \textit{Hypertensive heart disease} (I11; GR113-056) and \textit{Heart failure} (I50; GR113-067), among other circulatory diseases.

\begin{figure}[!h]
    \centering
    \includegraphics[width=\textwidth]{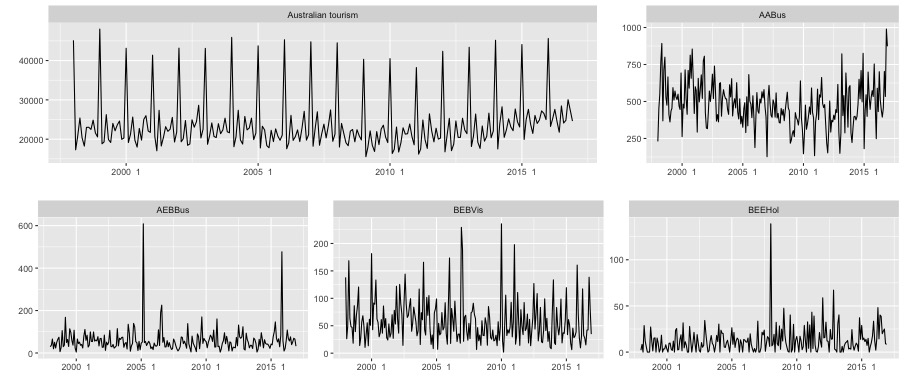}
    \caption{Visualization of selected time series from the tourism dataset. ``AA'', ``AEB'', ``BEB'', and ``BEE'' represent the zone ``Metro NSW'', and the regions ``New England North West'', ``Western Grampians'', and ``Spa Country'', respectively. ``Bus'', ``Vis'', and ``Hol'' denote travel purposes ``Business'', ``Visit'', and ``Holiday'', respectively.}
    \label{fig:tourism}
\end{figure}

\begin{figure}[!h]
    \centering
    \vspace{0.2in}
    \includegraphics[width=\textwidth]{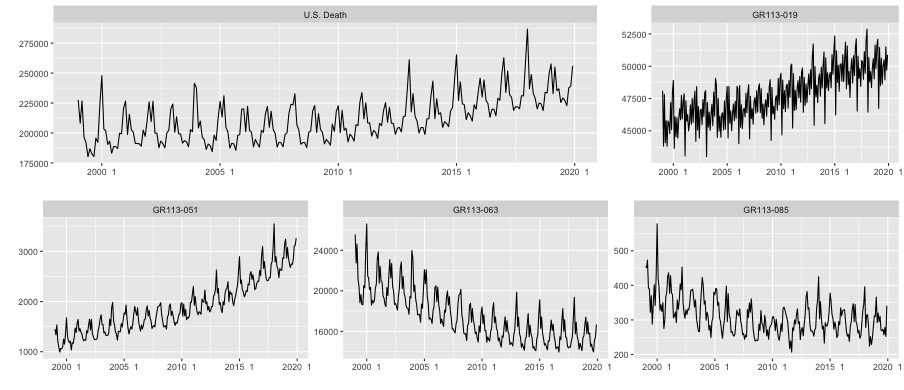}
    \caption{{Visualization of selected time series from the death count dataset.} ``GR113-019'', ``GR113-051'',  ``GR113-063'', and ``GR113-085'' denote ``Malignant neoplasms'', ``Parkinson disease'', ``All other forms of chronic ischemic heart disease'', and ``Asthma'', respectively.}
    \label{fig:mortality}
\end{figure}

The top-level series, one selected middle-level series, and three selected bottom-level series are illustrated in Figures~\ref{fig:tourism} and~\ref{fig:mortality} for the tourism and mortality datasets, respectively. 
Both figures exhibit more regular and apparent trend and seasonality for series at a higher level of aggregation, while bottom-level series are noisier and prone to outliers. 
Comparing the datasets to one another, the mortality dataset generally exhibits stronger seasonality and trend, whereas the tourism dataset displays greater volatility. Table~\ref{tab:features} summarises three features of each dataset, with larger values indicating more ``signal'' relative to ``noise''. These are: the Holt Winters seasonal smoothing parameter; the lag 12 autocorrelation coefficient; and the strength of trend measured as the proportion of variance explained by the trend component in an STL decomposition. Table~\ref{tab:features} supports the conclusions made by visualizing the time series of the data, which is that the tourism data are noisier and less regular with respect to trend and seasonality.\\

\begin{table}[!h]
    \centering
    \caption{\label{tab:features}Trend and seasonality features for the tourism  and mortality dataset.}
    \begin{tabular}{lrr}\toprule
        Feature & tourism &  mortality\\ \midrule
        Strength of trend & 0.1559 & 0.7574 \\
        Seasonality smoothing parameter & 0.0002 & 0.0202 \\ 
        Seasonal auto-correlation coefficient &0.1814  &0.6523  \\ 
 \bottomrule
    \end{tabular}
\end{table}

\subsection{Evaluation of forecast accuracy}
\label{subsec:evaluation}

To measure forecast accuracy, we first calculate the Root Mean Squared Scaled Error (RMSSE, \citealp{makridakisM5AccuracyCompetition2022}), for each series. {We define RMSSE of the $h$-step-ahead forecasts, produced using the $j$-th method and conditioning on $T$ observations, of the $i$-th time series as

\[
RMSSE_{j, i} = \sqrt{\frac{\frac{1}{h}\displaystyle\sum_{t=T+1}^{T+h}(y_{t, i}-\breve y_{t, j, i})^2}{\frac{1}{T-12}\displaystyle\sum_{t=13}^T (y_{t, i} - y_{t-12, i})^2}},
\]
where $\breve y_{t,j,i}$ be forecast of the $i$-th time series, generated by the $j$-th method, at time $t$.} RMSSE is symmetric, independent of the data scale, and thus suitable for evaluating hierarchical forecasts (\citealp{athanasopoulosEvaluationHierarchicalForecasts2023}). It should be noted that the denominator of our RMSSE measure slightly differs from that used in \cite{makridakisM5AccuracyCompetition2022}; we replace the naive forecast, with the seasonal naive forecast since the time series in our applications exhibit monthly seasonality. 
As a single measure of accuracy, we take the average RMSSE across all series. Since we will be comparing hierarchies with different structures (thus different middle-level series), this average only includes top and bottom level series, both of which are guaranteed to be present in all hierarchies.

We utilize the expanding window strategy to evaluate the performance of different approaches with a forecast horizon of 12 months (one year). For both datasets, the first window contains $96$ training observations. The training window is then increased by one observation and new 12-step ahead forecasts are obtained. This procedure is repeated until the training window comprises all but the  last $12$ observations. This results in $121$ $12$-step-ahead forecasts (January 2006 - January 2016) for the tourism dataset and $145$ $12$-step-head forecasts (January 2007 - January 2019) for the mortality dataset. 

To assess whether differences in forecast performance are statistically significant, we employ the Multiple Comparison with the Best (MCB) test (\citealp{koningM3CompetitionStatistical2005}). This test is based on the average ranks of different approaches across all evaluation windows and controls for multiple comparisons. { We implement the MCB test using the \code{tsutils} (\citealp{tsuitls}) R package}.

\subsection{Cluster hierarchies vs benchmarks}
\label{subsec:cluster_vs_benchmarks}

\begin{figure}[!h]
    \centering
    \vspace{0.1in}
    \includegraphics[width=0.45\textwidth]{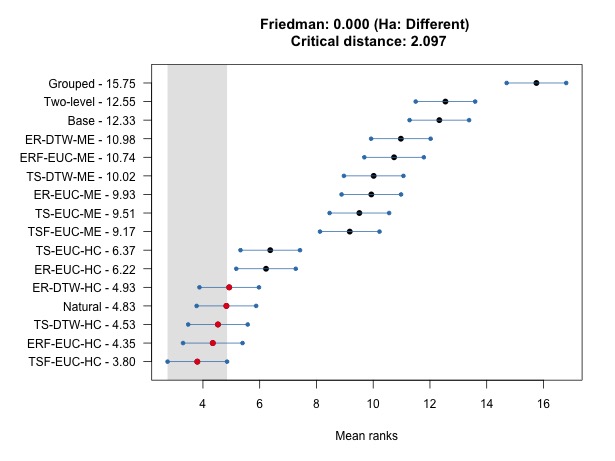}
    \includegraphics[width=0.45\textwidth]{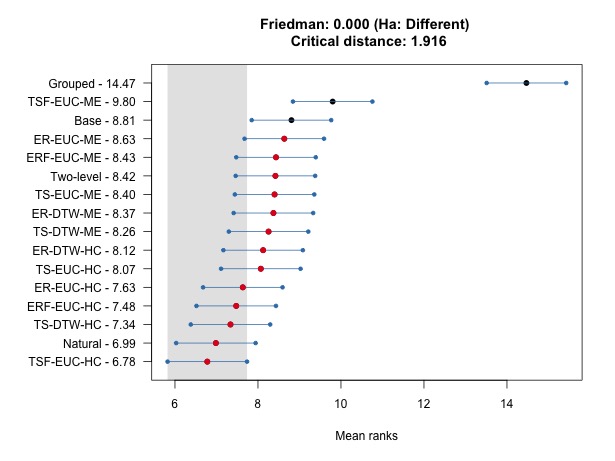}
    \caption{\label{fig:P3_mcb_benchmark}Average ranks and 95\% confidence intervals for { thirteen} cluster hierarchies and three benchmarks on tourism dataset (left) and mortality dataset (right) based on the MCB test.}
\end{figure}

Table~\ref{tab:P3_rmsse} compares the accuracy of reconciled forecasts when using hierarchies obtained from the $12$ clustering-based hierarchies outlined in Table~\ref{tab:P3_methods}. The base forecasts, as well as the reconciled forecasts from the two-level hierarchy (only containing top- and bottom- level time series) and from the natural hierarchy, are included as benchmarks.
The MCB test results are presented in Figure~\ref{fig:P3_mcb_benchmark}.

\begin{table}[!h]
    \centering
\caption{\label{tab:P3_rmsse}Performance of cluster hierarchies and benchmark hierarchies in terms of average RMSSE across all evaluation windows on both datasets. Column-wise minimum values are displayed in bold.}
\begin{tabular}{lrr}\toprule
    Approach & tourism & mortality \\ \midrule
    Base & 0.6945 & 0.7530 \\ 
    Two-level & 0.6944 & 0.7528 \\ 
    Natural & 0.6913 & 0.7501 \\ 
    Grouped & 0.7068 & 0.8091\\
    TS-EUC-ME & 0.6939 & 0.7528 \\ 
    ER-EUC-ME & 0.6938 & 0.7530 \\ 
    TSF-EUC-ME & 0.6938 & 0.7549 \\ 
    ERF-EUC-ME & 0.6942 & 0.7532 \\ 
    TS-EUC-HC & 0.6922 & 0.7540 \\ 
    ER-EUC-HC & 0.6920 & 0.7507 \\ 
    \textbf{TSF-EUC-HC} & \textbf{0.6909} & 0.7509 \\ 
    ERF-EUC-HC & 0.6910 & 0.7501 \\ 
    TS-DTW-ME & 0.6940 & 0.7528 \\ 
    \textbf{TS-DTW-HC} & 0.6911 & \textbf{0.7496} \\ 
    ER-DTW-ME & 0.6942 & 0.7531 \\ 
    ER-DTW-HC & 0.6912 & 0.7532 \\ \bottomrule
\end{tabular}

\end{table}

We have the following observations from Table~\ref{tab:P3_rmsse} and Figure \ref{fig:P3_mcb_benchmark}. In terms of average RMSSE, for both datasets, the base forecasts provide the worst forecast performance. The natural hierarchies provide better results than the base forecasts and the two-level hierarchies, and comparable results with cluster hierarchies. In the case of the tourism dataset, all twelve clustering-based hierarchies outperform the simple two-level hierarchy. For ten out of twelve of these methods, the prediction intervals for the average ranks do not overlap with the two-level hierarchy, indicating significantly superior performance.
For the mortality dataset, five cluster hierarchies surpass the two-level hierarchy in terms of average RMSSE. However,  not even the best clustering method is significantly more accurate than the two-level hierarchy based on the MCB test. {It should also be noted that the performance of the grouped hierarchy is much poorer compared to individual clustering hierarchies across both datasets. We suspect this may be a consequence of having too large a hierarchy, resulting in difficulty in estimating the covariance matrix $\boldsymbol{W}$ used as an input to MinT (\citealp{pritulargaStochasticCoherencyForecast2021}).}

The varying performance of cluster hierarchies across two datasets can be attributed to the unique characteristics of their bottom-level series.
The tourism dataset, as shown in Figure~\ref{fig:tourism}, predominately contains volatile and noisy bottom-level time series with weak trend and seasonality (see also in  Table~\ref{tab:features}). Arguably, creating new middle-level time series in this context helps elucidate the underlying pattern which can not be easily captured by bottom-level base forecasting models due to a low signal to noise ratio. 
On the other hand, the bottom-level series in mortality dataset exhibit stronger trend and seasonality patterns, meaning that the addition of middle-level series is less beneficial.\\

\begin{table}[!h]
    \centering
    \caption{\label{tab:P3_number_series}Average number of middle-level time series resulting from $k$-medoids clustering, hierarchical clustering, and the natural hierarchy for both datasets.}
    \begin{tabular}{lrr}
    \toprule
        Approach &  tourism& mortality \\ \midrule
     Natural &250  & 21 \\ 
        $k$-medoids clustering &  21&  7\\ 
        Hierarchical clustering &302   & 96\\ \bottomrule
    \end{tabular}
\end{table}
%20.99&  6.78

We also observe that the hierarchies constructed via hierarchical clustering algorithms outperform the hierarchies based on $k$-medoids when using the same representation and distance metric. As an example, ``TSF-EUC-HC'' outperforms ``TSF-EUC-ME''.
This superior performance can be attributed to hierarchical clustering generating a greater number of middle-level time series than $k$-medoids.
Table~\ref{tab:P3_number_series} summarizes the average number of middle-level series across all evaluation windows for natural, $k$-medoids, and hierarchical clustering hierarchies. Regarding the superiority of any specific representation or distance metric, no consistent findings emerge. It shows that while it is possible to improve forecast accuracy via clustering, the performance of different clustering methods largely depends on the specific data in consideration.
Interestingly, the natural hierarchy shows competitive accuracy compared to hierarchical clustering methods on both datasets, despite having fewer middle-level series. { The competitiveness of the natural hierarchy may be explained by factors other than similar series being combined together. For example, the natural hierarchy encodes prior information about the series, regarding common underlying shocks and dynamics for neighboring regions or similar causes of death. These may be more clearly detected and modeled in aggregate series.} However, it should be noted that natural hierarchies may not always exist. { For completeness, we further evaluate the forecast accuracy of middle-level time series, defined according to the natural hierarchy. For clustering-based hierarchies, we use a bottom-up approach. The results are consistent with our findings and are provided in the supplementary materials. }

%\section{Investigating driving factors: grouping and structure} 

\section{Disentangling  grouping and structure}
\label{sec:permutation}

These results in Section~\ref{subsec:cluster_vs_benchmarks} raise the question of why hierarchies augmented with middle levels improve upon the two-level hierarchy. There are two possible explanations.  On the one hand, it could be argued that by ``grouping'' together series with similar characteristics, certain signals are enhanced, leading to improved forecasting performance.
Alternatively, it could be argued that the improved forecasting performance is a by-product of the ``structure'' of the hierarchy, \textit{i.e.} the number of middle-level series, the depth of the hierarchy, the distribution of group sizes in the middle layer(s), or a combination of all these factors.
\subsection{Permutation hierarchy construction}
\label{subsec:permutation}

\begin{figure}[!h]
    \centering
    \includegraphics[width=0.9\textwidth]{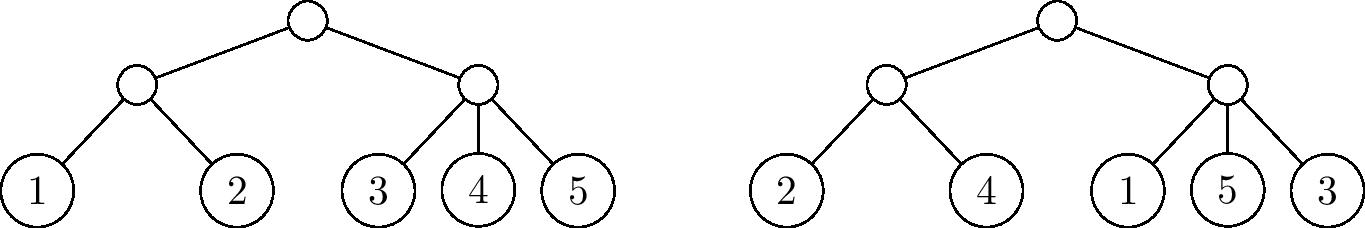}
\caption{\label{fig:aggcluster_random}Examples of a given hierarchy and its ``twin''. }
\end{figure}

To assess whether ``grouping'' or ``structure'' has relatively more importance, we consider a procedure based on permutation. For ease of exposition, we will describe this procedure for the natural hierarchy, although it is applied equally to hierarchies where middle-level series are constructed using clustering algorithms. First, the structure of the natural hierarchy is kept fixed. A new ``twin'' hierarchy is then constructed by randomly permuting the bottom level series\footnote{This can also be achieved by shuffling the columns of $\boldsymbol{C}$.}. An example is shown in Figure~\ref{fig:aggcluster_random}. { In this figure, we begin with the hierarchy on the left hand side and shuffle the leaves (the nodes on the bottom labelled 1--5) to obtain the hierarchy on the right hand side. Note that none of the middle levels from the hierarchy on the left are retained. Instead, where the hierarchy on the left has a middle-level variable formed by adding node 1 and node 2, the corresponding node on the right would have a middle level that adds node 2 and node 4. %These middle level nodes are both shaded grey in the figure. 
If nodes are aggregated on the basis of similarity, this will no longer hold in the permuted hierarchy. For example, in the left hand side hierarchy, node 1 and node 2 may correspond to similar time series. However, after permutation, the same structure may no longer add similar time series (\textit{i.e.} node 2 and node 4 may not be similar). In this way, permutation retains the structure of a hierarchy while removing the effect of grouping similar time series together.}

Suppose the forecast performance of the natural hierarchy can be explained by ``grouping''. In this case, the ``twin'' hierarchy, with a grouping formed at random, should perform significantly worse than the natural hierarchy. Alternatively, suppose the ``twin'' performs similarly to the natural hierarchy. In this case, the critical factor in improved forecast performance is the structure of the hierarchy, which is the same for both the natural hierarchy and its twin. To rule out the possibility that a random twin with exceptionally good (or bad) grouping is generated by chance, in all cases we consider 100 random twin hierarchies.

\subsection{Natural hierarchy vs its twins}

\label{subsec:n_vs_pn}

\begin{figure}[!h]
    \centering
    \includegraphics[width=0.47\textwidth]{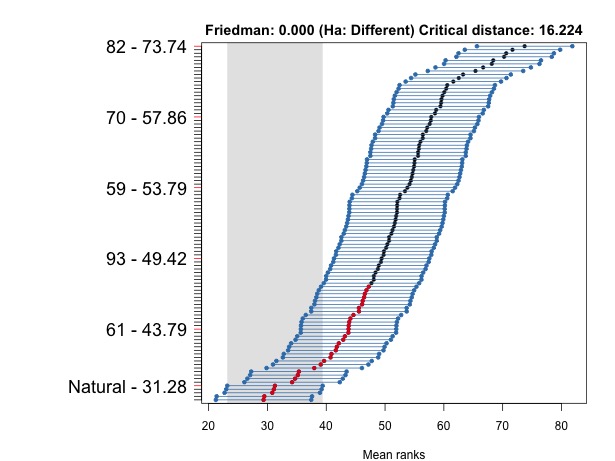}
    \centering    \includegraphics[width=0.47\textwidth]{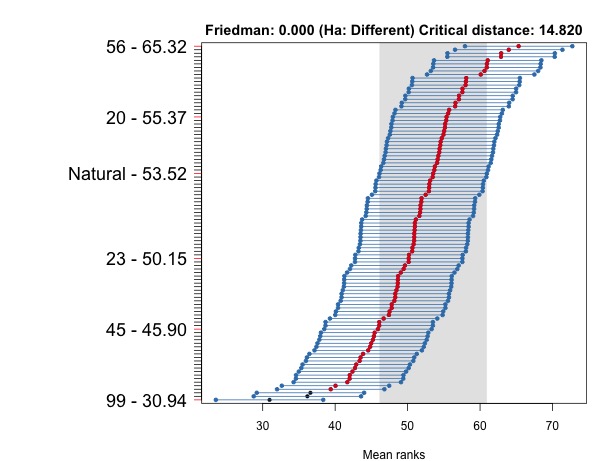}
    \caption{\label{fig:P2_tourism}Average ranks and 95\% confidence intervals for natural hierarchy and its $100$ twins, tourism dataset (left) and mortality dataset(right).}
\end{figure}

Figure~\ref{fig:P2_tourism} compares the natural hierarchy to 100 twins using the MCB test. For brevity, we only display the average rank labels for the natural hierarchy and $5$ of its twin hierarchies\footnote{Note that ``82 - 73.74'' represents that the $82$nd permutation of the hierarchy which has an average rank of 73.74. The same convention applies to the labels of the $y$-axis for Figures \ref{fig:P3_tourism_c_vs_pc},  \ref{fig:simu_P3_benchmarks}, and \ref{fig:P4_a_vs_pa}.}. The figure is adjusted so that the grey band is around the 95\% confidence interval for the natural hierarchy rather than the best performing twin. Any hierarchies whose confidence interval overlaps with the grey zone is not significantly better or worse than the natural hierarchy. 

It is clear that for both datasets, the natural hierarchy does not significantly outperform a large proportion of its random twins. 
On the tourism dataset, the natural hierarchy ranks the $5$th. Its performance is statistically indistinguishable from 32 of its twins, but significantly better than the remaining 68. The difference in forecast performance for the natural hierarchy and its twins is even less pronounced for the mortality dataset. As shown in right panel of Figure~\ref{fig:P2_tourism}, the performance of the natural hierarchy is statistically indistinguishable from most of its twins and, there are three twin hierarchies significantly better than the natural hierarchy. 
In both datasets, and particularly for the mortality dataset, we conclude that the ``structure'' of the natural hierarchy is the primary contributor to the improvement in forecast accuracy over the two-level hierarchy.

\subsection{Cluster hierarchy vs its twins}
\label{subsec:P3_c_vs_pc}

The result in Section \ref{subsec:n_vs_pn} suggesting that ``structure'' is a more important contributor to ``grouping'' may arise since the grouping for the natural hierarchy is not selected in a data-driven fashion. 
To assess whether clustering methods select a better ``grouping'', we compare the best-performing clustering-based hierarchies (TSF-EUC-HC and TS-DTW-HC for tourism and mortality, respectively) with their random twins.  Recall that Section~\ref{subsec:cluster_vs_benchmarks} demonstrates that the clustering methods can outperform the natural hierarchy and two-level hierarchy. 

\begin{figure}[!h]
    \centering
    \includegraphics[width=0.47\textwidth]{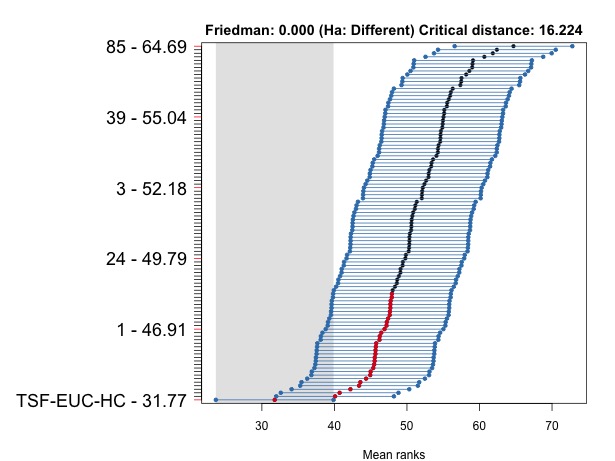}
    \includegraphics[width=0.47\textwidth]{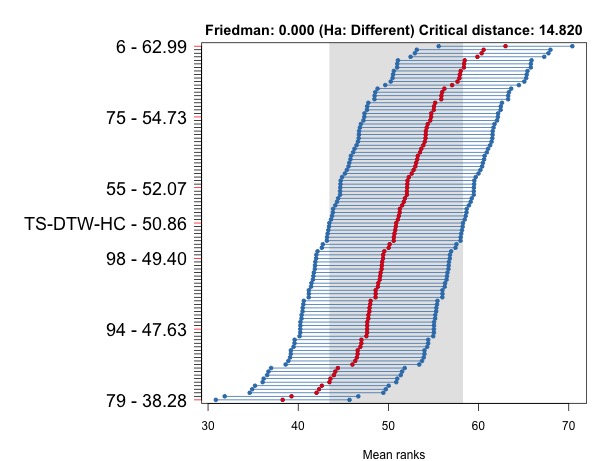}
    \caption{\label{fig:P3_tourism_c_vs_pc}Average ranks and 95\% confidence intervals for best performing clustering approach and its $100$ twins, tourism dataset (left) and mortality dataset (right).}
\end{figure}

The MCB test results for the tourism dataset and mortality dataset are shown in Figure~\ref{fig:P3_tourism_c_vs_pc}.
We observe that in both datasets, the best performing clustering approach does not yield significantly better results than its random twins. 
The best cluster approach of mortality dataset ranks nearly in the middle of its random twins, indicating that once again structure rather than grouping is the main driver of improvement in forecast accuracy. 
On the other hand, the best performing clustering method for the tourism data is statistically indistinguishable from 30 of its random twins, despite being the best in terms of the mean ranks. One can argue that for tourism dataset, a data-driven method for grouping time series plays a more prominent role in forecast improvement. 
This may be attributed to the noisier nature of bottom level tourism data, suggesting that similar weak signals are strengthened when aggregated.
However, there is still roughly a 30\% chance that a random twin performs similarly, once again highlighting that structure is the main contributor to improved forecast performance.

\section{Simulation study}
\label{sec:simulation}
The main conclusion of Section~\ref{sec:permutation} is that hierarchies with more middle-level series improve forecast accuracy due to structure rather than grouping. In so far as grouping may be a factor, this may be due to aggregating similar weak signals into stronger signals. To further test this conjecture, we consider a simulation study. Time series are generated that form clear clusters according to their trend and seasonality. These are then aggregated into middle-level series, based on the known characteristics of each time series. The purpose of this is two-fold. First, in a simulated setting, the ``true'' clusters can be known, which guards against the risk that a given clustering algorithm fails to identify the correct clusters. Second, a simulation study guards against the shortcoming of any cluster analysis, namely that clusters will always be found even where they are not present. Such spurious clusters may explain the similar forecasting performance of cluster-based hierarchies with their randomly permuted twins. The simulation study thus sets up an ideal scenario, where grouping can potentially dominate structure as the factor explaining improved forecast performance.

\subsection{Simulation design}

We construct $m=120$ bottom-level time series in an additive manner, each following a data generating process described as follows:
\begin{equation}
    \label{simu:DGP}
    \begin{aligned}
    Y_t &= L_t + S_t + \xi_t, \\
        L_t &= \alpha t + \varepsilon_t,\\
    S_t &= \begin{cases}\beta\quad\textrm{if $t-\delta$ is even}\\\gamma\quad\textrm{if $t-\delta$ is odd}\end{cases}\,,
    \end{aligned}
\end{equation}
where $L_t$ represents the trend term that increases or decreases over time with slope $\alpha$. We set $\alpha$ to 0.001, $-0.002$, and 0 so that exactly one third of the bottom level series have increasing, decreasing, and no trend respectively. The seasonal pattern is determined by $S_t$. It is deterministic with a seasonal period of 2, hitting a peak $\beta$ drawn uniformly from [2, 3], and a trough $\gamma$ drawn uniformly from [0, 1]. The parameter $\delta$ controls whether a time series has its seasonal peak for odd values of $t$ or even values of $t$, which we refer to as ``odd'' and ``even'' seasonality respectively. We set $\delta=0$ (even seasonality) for exactly half of the series and $\delta=0$ (odd seasonality) for the other half of the series. Both $\xi_t$ and $\varepsilon_t$ are white noise with the variance of $\xi_t$ set to 0.25 and the variance of $\varepsilon_t$ set to $2.5\times 10^{-5}$ and $4.9\times 10^{-5}$ for increasing trend and decreasing trend, respectively. The combination of three different trends with two different patterns of seasonality leads to six clusters as described in Table~\ref{table:simu_params}.

\begin{table}[!h]
\caption{\label{table:simu_params}Parameter setting for all clusters in the simulation experiments.}
\centering
\begin{tabular}{lcccccc}\toprule
& Cluster 1 & Cluster 2 & Cluster 3 & Cluster 4 & Cluster 5 & Cluster 6 \\ \midrule
Trend & Increase & Increase & None & None & Decrease & Decrease \\
Seasonality & Odd & Even & Odd & Even & Odd & Even  \\
    \bottomrule
\end{tabular}
\end{table}

\begin{figure}[!h]
\centering
\includegraphics[width=\textwidth]{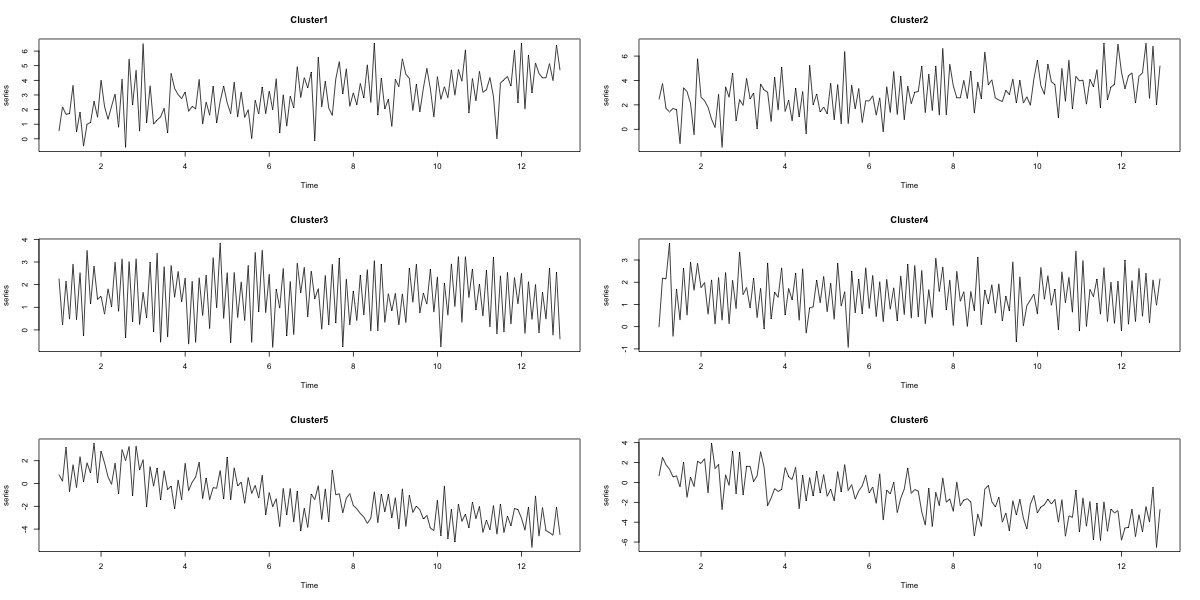}
\caption{\label{fig:simu_emps}Example time series for each cluster in the simulation experiments.}
\end{figure}

\begin{figure}[!h]
    \centering
    \includegraphics[width=0.65\textwidth]{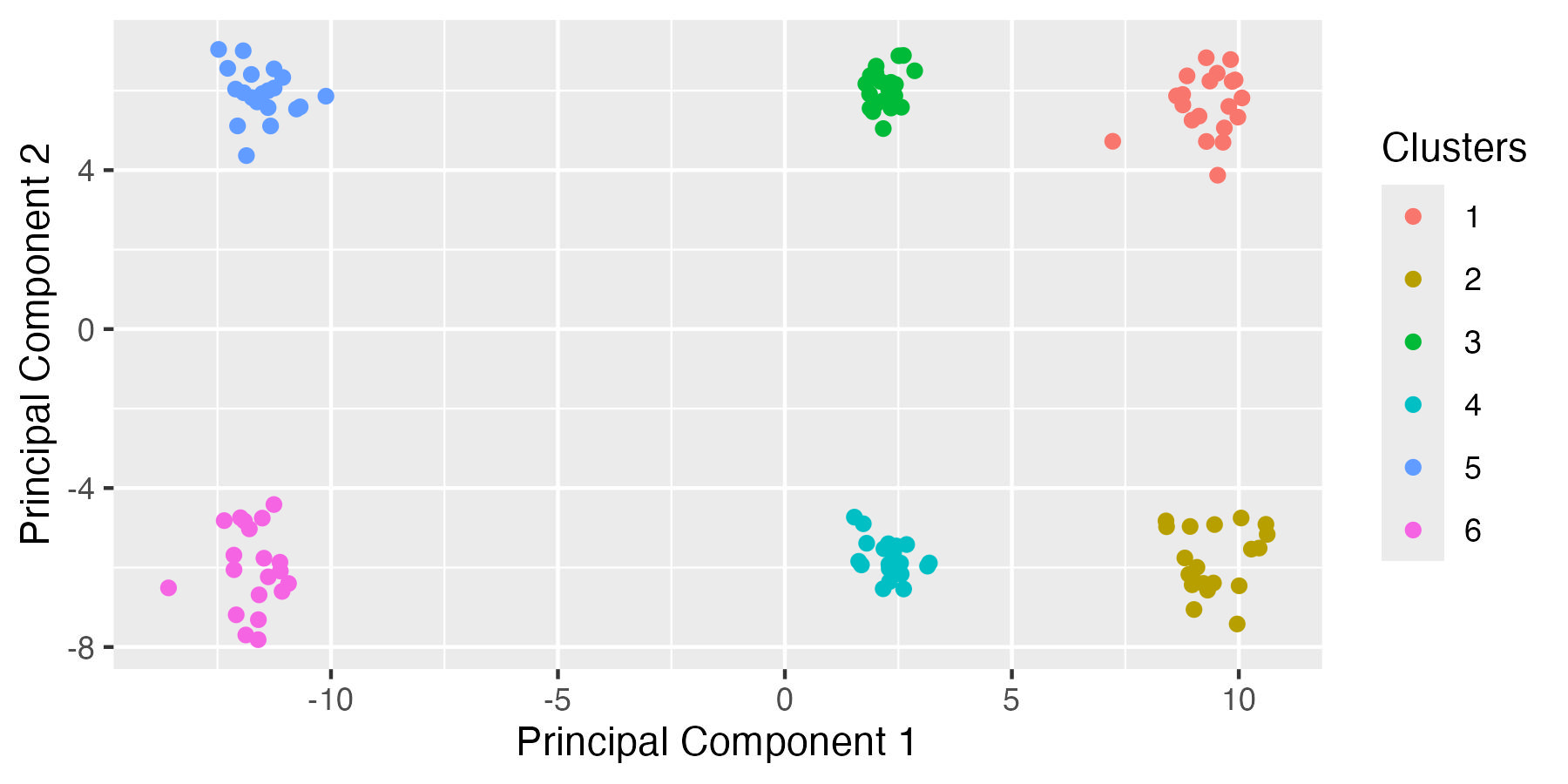}
    \caption{\label{fig:simu_pca}Visualization of the generated time series in the simulation experiments.}
\end{figure}

For each series, we generate $144$ observations, with the last $12$ observations reserved for evaluation. Figure~\ref{fig:simu_emps} displays typical time series from each cluster, while Figure~\ref{fig:simu_pca} is a scatterplot of the first two principal components of the series. It is clear that the simulation design generates distinct clusters.

Various schemes for constructing middle-level series can be considered and are summarised in Table~\ref{tab:7}. middle-level series can be constructed according to the value of the $\alpha$ parameter, leading to three clusters, or according to whether the series has a trend or no trend, leading to two clusters. In Table~\ref{tab:7}, these are referred to as Cluster-trend1 and Cluster-trend2 respectively. middle-level series can also be formed on the basis of seasonality (Cluster-season), leading to two clusters, or according to both trend and seasonality leading to six clusters (Cluster-trend-season).

\begin{table}[!h]
    \centering
    \caption{\label{tab:simu_methods}Four clustering approaches used in the simulation experiments.}
    \begin{tabular}{ll}\toprule
        Approach & Description \\ \midrule
        Cluster-trend-season & Hierarchy based on trend and seasonal pattern. \\
        Cluster-trend1 &  Hierarchy based on trend (positive/negative/none). \\
        Cluster-trend2 & Hierarchy based on trend/no trend. \\
        Cluster-season & Hierarchy based on seasonal pattern (odd/even). \\\bottomrule
    \end{tabular}
    \label{tab:7}
\end{table}

\subsection{Results}
\label{sec:simu_res}

\begin{figure}[!h]
    \centering
    \vspace{0.1in}\includegraphics[width=0.7\textwidth]{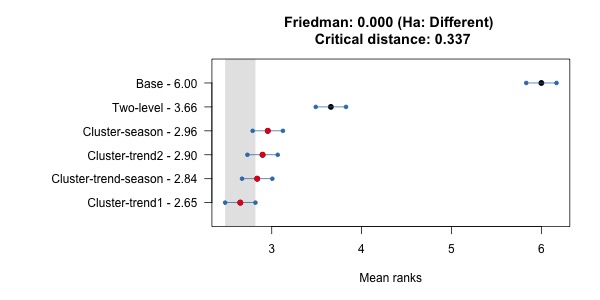}
   \vspace{-0.1in}
\caption{\label{fig:simu_P3_benchmarks}Average ranks and 95\% confidence intervals for four cluster hierarchies and two benchmarks in simulation based on MCB test.}
\end{figure}

We replicate the simulation $500$ times and follow the evaluation procedure introduced in Section~\ref{subsec:evaluation}.
Table~\ref{tab:simu_P3} reports the average RMSSE across all hierarchies, and
Figure~\ref{fig:simu_P3_benchmarks} presents the MCB test results. The results reveal that most approaches perform better than the base forecasts and the two-level hierarchy. This outcome indicates that hierarchy construction generally improves forecast reconciliation performance, corroborating our findings reported in Section~\ref{subsec:cluster_vs_benchmarks}. 
Clustering only on the basis of trend yields the best performance, however all clustering schemes are statistically indistinguishable from one another.

\begin{table}[!h]
    \centering
    \caption{\label{tab:simu_P3}Performance of cluster hierarchies and benchmark hierarchies in terms of average RMSSE in simulation. Column-wise minimum values are displayed in bold.}
    \begin{tabular}{lc}\toprule
        Approach & Average RMSSE \\ \midrule
        Base & 0.7764 \\ 
        Two-level & 0.5971 \\ 
        Cluster-trend-season & 0.5963 \\ 
        Cluster-trend1 & \textbf{0.5962} \\ 
        Cluster-trend2 & 0.5965 \\ 
        Cluster-season & 0.5965 \\ \bottomrule
    \end{tabular}
\end{table}

\begin{figure}[!h]
    \centering
\includegraphics[width=0.8\textwidth]{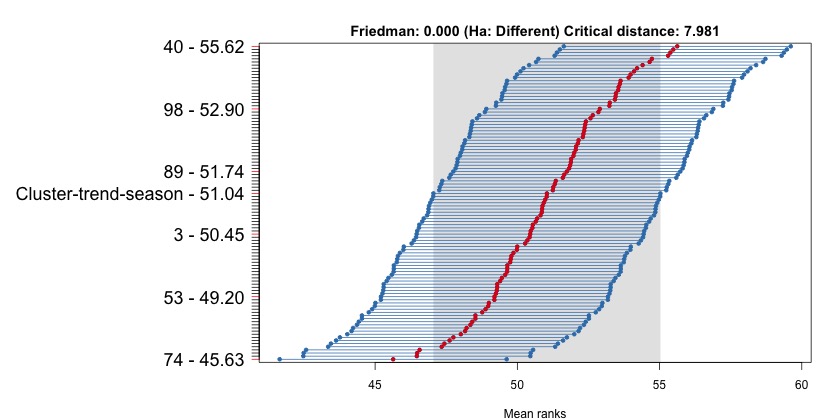}
    \vspace{-0.1in}
    \caption{Average ranks and 95\% confidence intervals for four the ideal hierarchy and its 100 random twins in simulation based on MCB test.}
    \label{fig:simu_P3_c_vs_pc}
\end{figure}

Having constructed clearly demarcated clusters, we now compare the six cluster hierarchy against 100 of its random twins.  The MCB test result is shown in Figure~\ref{fig:simu_P3_c_vs_pc}. Using the known clusters yields a performance somewhere in the middle of the 100 twins that is statistically indistinguishable from nearly all twins. %\todo{is it really all of them? Yes. In the MCB plot, all elements indistinguishable from the target element (here the Cluster-predefined) have red points}. 
This arguably provides the most compelling evidence so far on whether improvements in forecast accuracy can be attributed to structure or grouping. It is the process of forming any middle-level series that is more important than aggregating similar time series. This result holds not only for our empirical studies, but also for an example specifically designed to have distinct, known clusters as illustrated in this section.

\section{Forecast combination}
\label{sec:combination}

The results in Sections~\ref{sec:cluster} and~\ref{sec:simulation} highlight the potential of improving forecast reconciliation performance through the construction of new middle levels series using time series clustering. The selection of the best performing combination of time series representation, distance measure, and clustering algorithm remains an open question. Since the best cluster-based method will heavily depend on the characteristics of the data, %and forecasting method used to obtain bottom level series, 
we consider averaging forecasts across different hierarchies, as an alternative to hierarchy selection. This is supported by substantial research and empirical evidence in favor of forecast combination over selection (see \textit{e.g.}, \citealp{elliottForecastingEconomicsFinance2016}). Specifically, the reconciled forecasts from multiple hierarchies are combined using equal weights (\citealp{wangForecastCombinations50year2022}), \textit{i.e.},
\[
  \tilde{\boldsymbol{y}}_{T+h}^{\text{comb}} = \frac{1}{l} \sum_{j=1}^l \tilde{\boldsymbol{y}}_{T+h}^j,
\] 
{where $l$ is the number of hierarchies considered in the forecast combination analysis.}
We note that this average can only be carried out using series that are common to all hierarchies, in our case, the top and bottom levels. In this case, since all elements of the average are coherent, any linear combination of these forecasts will also be coherent. \\

\begin{table}[!h]
    \centering
    \caption{\label{tab:P4_RMSSE}Average RMSSE across six approaches. Column-wise minimum values are displayed in bold.}
    \begin{tabular}{lcc}\toprule
       Approach & tourism & mortality \\ \midrule
        Base & 0.6945 & 0.7530 \\ 
        Two-level & 0.6944 & 0.7528 \\ 
        Natural & 0.6913 & 0.7501 \\ 
        TS-DTW-HC & 0.6911 & 0.7496 \\ 
        TSF-EUC-HC & 0.6909 & 0.7509 \\ 
        Combination & \textbf{0.6902} & \textbf{0.7423} \\ \bottomrule
    \end{tabular}
\end{table}

\begin{figure}[!h]
    \centering
    \includegraphics[width=0.45\textwidth]{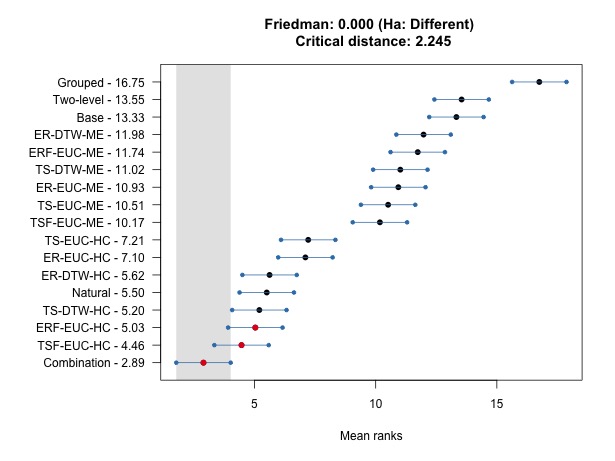}
    \includegraphics[width=0.45\textwidth]{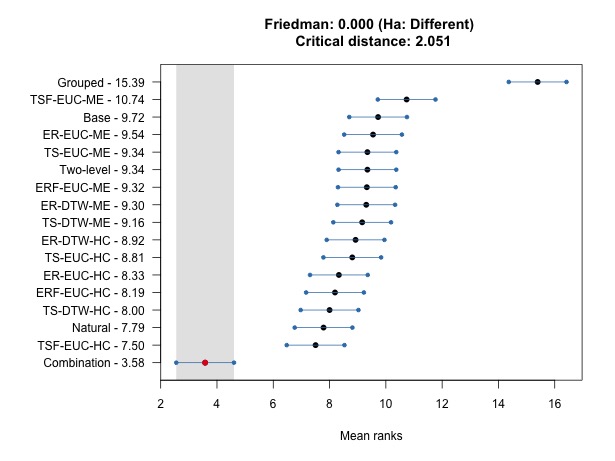}
    \caption{\label{fig:P4_bench_mcb}Average ranks and 95\% confidence intervals for all approaches on tourism dataset(left) and mortality dataset(right) based on MCB test.}
\end{figure}

Table~\ref{tab:P4_RMSSE} presents the accuracy in terms of average RMSSE across all evaluation windows for both datasets. Note that for brevity we only present results for three benchmarks (base, two-level, and natural), the best cluster hierarchy (TS-DTW-HC for the mortality data and TSF-EUC-HC for the tourism data) and the combination hierarchy. The MCB test results are shown in Figure~\ref{fig:P4_bench_mcb}. 
As expected, we observe that on both datasets, forecast combination improves forecast performance compared to any single hierarchy. 
The improvement on the mortality dataset is more pronounced, with forecast combination significantly outperforming all other approaches.

\subsection{Combination hierarchy vs its random twins}

We also consider how the question of grouping versus structure plays out in the case of combinations by extending the permutation approach introduced in Section~\ref{subsec:permutation}. We do so by first permuting the labels of the bottom level series while keeping the hierarchy fixed. The same permutation is used for all hierarchies in a combination. This is then repeated 100 times, yielding 100 ``twins'' of the combination forecast.

\begin{figure}[!h]
    \centering
    \includegraphics[width=0.6\textwidth]{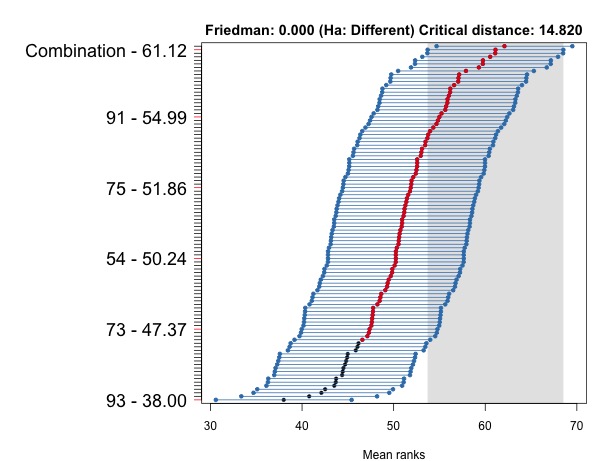}
    \vspace{-0.1in}\caption{\label{fig:P4_a_vs_pa} Average ranks and 95\% confidence intervals for combination of twelve cluster hierarchies and its $100$ random twins on mortality dataset based on MCB test.}
\end{figure}

The results of MCB test for the combination and its $100$ random twins are presented in Figure~\ref{fig:P4_a_vs_pa} for the mortality dataset\footnote{Note that this experiment has been conducted only on the mortality dataset, as computing random twins for all cluster hierarchies on the tourism dataset is too computationally expensive.}.
Similar to the results in Section~\ref{sec:permutation}, we find evidence that a large number of random twins do not perform significantly worse than an approach based on clustering. In fact, out of the 100 twins, the combination only performs better than a single twin. This provides the last piece of compelling evidence that the grouping of similar time series, contributes less to improved forecast performance than structure.

\section{Conclusion}
\label{sec:conclusion}

This paper thoroughly investigates the issue of constructing hierarchies for forecast reconciliation, with the goal of improving forecast accuracy. This issue is particularly important in the absence of a predefined natural hierarchy. Rather than focus on a single clustering algorithm, we consider a more general framework, incorporating three distinct approaches: cluster hierarchies, permutation hierarchies, and combination hierarchies. Unsurprisingly, no single method emerges as superior for all datasets, although hierarchical clustering, which by construction leads to a larger number of clusters, tends to outperform $k$-mediods clustering.

This naturally begs the question, of why adding more middle-level series can improve forecast performance. We devise a method that keeps the hierarchy fixed while permuting the bottom level series, yielding ``twin'' hierarchies. The bottom-level series that aggregate to a single middle-level series are thus chosen at random. Across different datasets and settings, an overwhelmingly large number of twins, perform not significantly worse, or even better, than a hierarchy selected using clustering. This suggests that the improved forecast performance arises not from grouping similar times series together, but rather from features related to the structure of the hierarchy, \textit{e.g.} the larger number of middle-level series. We find similar results from a simulation study where clusters are predefined. 
This eliminates the possibility that our empirical results arise due to  spurious clusters or the inadequate performance of a single clustering method.
 
Our main practical recommendation is to use multiple clustering methods and combine forecasts across these methods using equal weights combination. 
This mitigates the uncertainty of selecting the best clustering approach and is shown to significantly outperform all benchmarks across both datasets that we consider. %One could also extend this idea to averaging over random twins, although we note that any averaging approach incurs a computation cost in needing to obtain base forecasts for more middle-level series.
One could also extend this idea to averaging over random twins. However, it's worth noting that any averaging approach incurs a computation cost as it requires obtaining base forecasts for additional middle-level series.

Future research based on our study could proceed in several promising directions. 
For example, while we used equally-weighted combinations in this study, there is  potential to apply more sophisticated forecast combination methods to improve performance. The extensive literature on forecast combination, including advanced methods for calculating weights, could also be considered (refer to \citealp{wangForecastCombinations50year2022} for an in-depth review).

Also, our results are based on cross-sectional data, this could be extended to explore temporal (\citealp{athanasopoulosForecastingTemporalHierarchies2017}) and cross-temporal hierarchies (\citealp{girolimettoCrosstemporalProbabilisticForecast2023a}). Finally, more work could be carried out to understand whether some middle levels contribute more to forecast accuracy than others, and accordingly ``pruning'' the less useful middle-level series.\\

\section*{Acknowledgements}

Bohan Zhang is supported by the Academic Excellence Foundation of Beihang University for PhD Students and international joint doctoral education fund of Beihang University, China. Han Li is supported by the Australian Research Council Discovery Projects funding scheme (DP220100090).

\section*{Declaration of competing interest}

The authors declare that they have no known competing financial interests or personal relationships that could have appeared to influence the work reported in this paper.

\begingroup
\setstretch{1.15}
\bibliographystyle{agsm}
\bibliography{references.bib}
\endgroup

\end{document}